\documentclass[onecolumn,showpacs,amsfonts,aps,prc,nofootinbib,floatfix,%
superscriptaddress]{revtex4}

\usepackage{amsmath}
\usepackage{bm}
\usepackage{graphicx}

\usepackage{epsfig}
\newcommand{\beq}{\begin{equation}}
\newcommand{\eeq}{\end{equation}}
\newcommand{\bea}{\vspace{0.25cm}\begin{eqnarray}}
\newcommand{\eea}{\end{eqnarray}}

\newcommand{\ro}{\mbox{{\boldmath
$\rho$}}}

\newcommand{\qb}{\mbox{{\bf
q}}}

\def\lsim{\mathrel{\rlap{\lower4pt\hbox{\hskip1pt$\sim$}}
    \raise1pt\hbox{$<$}}}         
\def\gsim{\mathrel{\rlap{\lower4pt\hbox{\hskip1pt$\sim$}}
    \raise1pt\hbox{$>$}}}         

\newcommand{\landau}{L.D.~Landau Institute for Theoretical Physics,
        GSP-1, 117940, Kosygina Str. 2, 117334 Moscow, Russia}

\begin{document}


\title{
Radiative parton energy loss in expanding 
quark-gluon plasma with magnetic monopoles
}

\author{B.G.~Zakharov}\affiliation{\landau}

\begin{abstract}
We study radiative parton energy loss in an expanding quark-gluon
plasma with magnetic monopoles.
We find that for realistic number density of thermal monopoles 
obtained in lattice simulations parton rescatterings on 
monopoles can considerably
enhance energy loss for plasma produced in $AA$ collisions at
RHIC and LHC energies. However, contrary to previous expectations,
monopoles do not lead to the surface dominance of energy loss.
\end{abstract}
%

\maketitle


\section{Introduction}

It is widely accepted that the jet quenching phenomenon in $AA$ collisions
observed at RHIC and LHC 
is a manifestation of parton energy loss in the hot quark-gluon
plasma (QGP) produced in the initial stage of $AA$ collisions. The
dominating contribution to parton energy loss comes from induced
gluon radiation due to parton multiple scattering in the QGP 
\cite{GW,BDMPS,LCPI,BSZ,W1,GLV1,AMY}.
The effect of collisional energy loss is relatively small \cite{Z_Ecoll}.
The RHIC and LHC data on suppression of the high-$p_{T}$ hadrons
in $AA$ collisions can be reasonably well described within
the light-cone path integral (LCPI) approach to induced gluon emission \cite{LCPI} in a scenario 
of purely perturbative QGP (pQGP) \cite{RAA08,RAA11,RAA12,RAA13} 
with the quasiparticle parton masses borrowed from the quasiparticle 
fit \cite{LH} to lattice results (which are close to that predicted by the HTL scheme \cite{HTL}). 
Although, in the relevant range of the plasma
temperatures $T\lsim T_{c}(2-3)$, the non-perturbative effects
may be important,
one could hope that the pQGP model is reasonable since radiative energy loss
is mostly sensitive to the number density of the color constituents of the QGP. 
And the internal dynamics of the matter is practically unimportant
from the standpoint of the energy loss calculations.
This assumption may be wrong, however, if the non-perturbative effects
lead to formation of new effective scattering objects that are absent in the
pQCD picture.
Evidently, thermal magnetic monopoles in the so called ``magnetic scenario'' of 
the QGP \cite{Shuryak_MC_mQGP,CZ1_mQGP,Shuryak_mQGP} are such objects
that can be potentially very important for parton energy loss. 

The thermal magnetic monopoles are now under active
investigation \cite{Born_M-dyon,Born_Rm1,Bonati,Mitr_mon} 
(and references therein). 
Lattice calculations  show that monopoles in the QGP are  
compact and heavy objects \cite{Born_Rm1}. 
For this reason from the point of view of parton
rescatterings they can act as practically point-like 
static scattering centers. Similarly to QED
(for a review on monopoles in QED see, for instance, 
\cite{Milton_rev}) the differential cross section for parton scattering
off thermal monopoles has the Rutherford form.  
It is important that, contrary to the ordinary pQCD parton cross sections,
for monopoles, due to the Dirac charge quantization condition constraint,
there are no the running and thermal effects. 
Lattice results
show that the monopole number density, $n_{m}$, 
may be quite large $n_{m}/T^{3}\sim 0.4-0.9$ at $T\sim T_{c}(1-3)$ 
\cite{Bonati,Born_Rm1}. Although, it is smaller
by a factor of $\sim 5-10$ than the number density of ordinary thermal partons
in pQGP, the scattering cross section for monopoles is considerably higher
than that for thermal quarks and gluons. 
As a result, the monopoles can give a considerable contribution to 
induced gluon  emission (and to photon emission from quarks).
In  \cite{Shuryak_m-phot} within the classical non-relativistic
approach it was shown that interaction of quarks with monopoles may be important
for photon emission from the QGP. The effect of monopoles on jet quenching
in $AA$ collisions has been addressed in recent analysis \cite{Gyulassy_mqgp}
within the GLV approach \cite{GLV1} in the approximation of $N=1$ rescattering.

In the present paper we address within the LCPI
scheme \cite{LCPI}  the question to which extent monopoles can be important for parton energy loss in the expanding
QGP for RHIC and LHC conditions. 
The advantage of the LCPI formalism is that it includes any 
number of parton rescatterings (that is very important for
the QGP with monopoles (below we denote it as mQGP) due to large cross section of parton interaction 
with monopoles). The LCPI approach treats accurately 
the mass and finite-size effects, and is valid beyond the soft
gluon approximation.

\section{Theoretical framework}
In the LCPI approach \cite{LCPI} the induced gluon $x$-spectrum for a fast quark (or gluon) 
 may be written
through the in-medium light-cone wave function 
of the $gq\bar{q}$ (or $ggg$) system in the coordinate $\rho$-representation.
The $z$-dependence of this light-cone wave function is governed by a 
two-dimensional Schr\"odinger equation 
in which the longitudinal coordinate $z$  ($z$-axis is chosen along the fast 
parton momentum) plays the role of time.
We use the representation for the gluon spectrum obtained in \cite{Z04_RAA}
which is convenient for numerical calculations. 
For a fast quark (produced at $z=0$) the gluon spectrum reads
\beq
\frac{d P}{d
x}=
\int\limits_{0}^{L}\! d z\,
n(z)
\frac{d
\sigma_{eff}^{BH}(x,z)}{dx}\,,
\label{eq:10}
\eeq
where $n(z)$ is the medium number density, $d\sigma^{BH}_{eff}/dx$ 
is an effective Bethe-Heitler
cross section accounting for both the LPM and finite-size effects.
The $d\sigma^{BH}_{eff}/dx$
reads 
\beq
\frac{d
\sigma_{eff}^{BH}(x,z)}{dx}=-\frac{P_{q}^{g}(x)}
{\pi M}\mbox{Im}
\int\limits_{0}^{z} d\xi \alpha_{s}(Q^{2}(\xi))
\left.\frac{\partial }{\partial \rho}
\left(\frac{\Psi(\xi,\rho)}{\sqrt{\rho}}\right)
\right|_{\rho=0}\,\,.
\label{eq:20}
\eeq
Here 
$P_{q}^{g}(x)=C_{F}[1+(1-x)^{2}]/x$ is the usual splitting
function for $q\to g q$ process,
$
M=Ex(1-x)\,
$
is the reduced "Schr\"odinger mass", $E$ is the initial parton energy,
$Q^{2}(\xi)=aM/\xi$ with $a\approx 1.85$ \cite{Z_Ecoll},
$\Psi$ is the solution to the radial Schr\"odinger 
equation for the azimuthal quantum number $m=1$ 
\beq
i\frac{\partial \Psi(\xi,\rho)}{\partial \xi}=
\left[-\frac{1}{2M}\left(\frac{\partial}{\partial \rho}\right)^{2}
+v(\rho,x,z-\xi)
+\frac{4m^{2}-1}{8M\rho^{2}}
+\frac{1}{L_{f}}
\right]\Psi(\xi,\rho)\,
\label{eq:30}
\eeq
with the boundary condition
$\Psi(\xi=0,\rho)=\sqrt{\rho}\sigma_{q\bar{q}g}(\rho,x,z)
\epsilon K_{1}(\epsilon \rho)$  
($K_{1}$ is the Bessel function),
$L_{f}=2M/\epsilon^{2}$
with $\epsilon^{2}=m_{q}^{2}x^{2}+m_{g}^{2}(1-x)^{2}$,
$\sigma_{q\bar{q}g}(\rho,x,z)$ is the cross section of interaction
of the $q\bar{q}g$ system (in the $\rho$-plane $\bar{q}$ is located at 
the center of mass of $qg$) with a medium constituent
located at $z$.
The potential $v$ in (\ref{eq:30}) reads
\beq
v(\rho,x,z)=-i\frac{n(z)\sigma_{q\bar{q}g}(\rho,x,z)}{2}
\label{eq:40}
\eeq
(summing over the species of the
medium constituents is implicit here).
The $\sigma_{q\bar{q}g}$ may be written through the dipole cross
section $\sigma_{q\bar{q}}$ 
\cite{NZ_sigma3}
\beq
\sigma_{q\bar{q}g}(\rho,x,z)=\frac{9}{8}
[\sigma_{q\bar{q}}(\rho,z)+
\sigma_{q\bar{q}}((1-x)\rho,z)]-
\frac{1}{8}\sigma_{q\bar{q}}(x\rho,z)\,.
\label{eq:50}
\eeq
%
The dipole cross section for scattering of the $q\bar{q}$ pair 
on a medium constituent $c$ may be written as
\beq
\sigma_{q\bar{q}}(\rho,z)=\frac{2}{\pi}\int d\qb
[1-\exp(i\qb\ro)]\frac{d\sigma_{qc}}{dq^{2}}\,,
\label{eq:60}
\eeq
where $d\sigma_{qc}/dq^{2}$ is the $qc\to qc$ differential cross section.
For scattering on thermal quarks and gluons 
the differential cross section (in the approximation of static Debye 
screened color centers
\cite{GW}) reads
\beq
\frac{d\sigma_{qc}}{dq^2}=\frac{C_{T}C_{F}}{2}
\frac{\pi \alpha_{s}^{2}(q^{2})}
{[q^{2}+m^{2}_{D}]^{2}}\,,
\label{eq:70}
\eeq
where $C_{F,T}$ are the color Casimir for the quark and thermal parton 
(quark or gluon), $m_{D}$ is the local Debye mass.
For the QGP with monopoles we should account for in the potential 
(\ref{eq:40}) the contribution from rescatterings on monopoles.
The formula (\ref{eq:60}) is valid for monopoles as well. 
In QCD there are two different species of monopoles related to 
the Cartan generators $T_{3}$ and $T_{8}$ of the $SU(3)$ group.
Lattice calculations \cite{Bonati}  show that both the species of 
monopoles have the same number density. For fast quarks thermal monopoles 
$M_{3,8}$ 
act as Abelian scattering centers. For gluons in the color basis 
of definite color isospin and hypercharge (see below) it is true as well.
In vacuum the differential cross section 
for scattering of a charged particle with electric charge $q_{e}$
off a monopole with magnetic charge $q_{m}$ has the Rutherford form
\cite{Urrutia}
\beq
\frac{d\sigma}{dq^2}=\frac{4\pi D^2 }{q^4}\,,
\label{eq:80}
\eeq   
where $D=q_{e}q_{m}/4\pi$.
The Dirac charge quantization condition says that $|D|=n/2$ where $n$ is
an arbitrary integer. We will assume that in the QGP for 
both the color species of monopoles $|D|=1/2$ (here we mean the minimal
value of $|D|$ for parton-monopole interactions, for some parton color states
it can be bigger).  
This value is supported by extraction of the magnetic coupling 
$\alpha_{m}=q_{m}^2/4\pi$ from the monopole-(anti)monopole
correlations in lattice simulations \cite{Born_Rm1,Bonati} which give
$\alpha_{m}\sim 2-4$ at $T/T_{c}\sim 1-2$. Making use this 
$\alpha_{m}$ by inspecting
the $qM$-scattering one can easily obtain that the condition $|D|=1/2$ 
gives $\alpha_{s}=1/\alpha_{m}\sim 0.25-0.5$ which is quite reasonable for
$\alpha_{s}$ in the QGP at $T/T_{c}\sim 1-2$. The value $|D|=1$ leads to
a four times bigger $\alpha_s$ which seems to be unrealistic.
For the $qM_{3,8}$ scattering 
the coupling to the vector potential of the monopole color field 
is given by $g\hat{\lambda}_{3,8}/2$, and the possible values of $|D|$ are 
$1/2$ and $1$.  
We write the $qM(\bar{M})$ differential
cross section in the form
\beq
\frac{d\sigma_{qM}}{dq^2}=\frac{C_{F}\pi F^2(q^2)}{(q^2+m_{D}'^{2})^{2}}\,.
\label{eq:90}
\eeq   
Here we introduced a  phenomenological form-factor $F$ accounting for 
the finite size of the monopole, $m_{D}'$ is the magnetic Debye mass,
$C_{F}$ is the color factor arising from averaging over
the quark and monopole color states which for the $SU(3)$ group gives
$[(1^{2}+(-1)^2)/3+(1^2+1^2+(-2)^{2})/3]/2=4/3$.
From (\ref{eq:70}) and (\ref{eq:90}) one can see that in comparison to $qq$ 
scattering the $qM(\bar{M})$  cross section is enhanced by a 
factor of $3/2\alpha_s^2$ (if we ignore $F$ and possible difference in 
the electric  and magnetic Debye screening masses). 

For energetic partons energy loss is dominated by the small $x$-region.
In the limit $x\to 0$ 
the three-body cross section (\ref{eq:50}) reduces to the cross
section for the $gg$-color dipole. In pQCD for scattering on thermal partons
$\sigma_{gg}=\frac{C_{A}}{C_{F}}\sigma_{q\bar{q}}$. One can show that this
relation is valid for scattering off monopoles as well.
Indeed, similarly to our analysis of the synchrotron-like gluon emission
\cite{Z_synch}, the scattering amplitude for interaction of gluons with 
monopoles may be  diagonalized by introducing the gluon fields having definite
color isospin, $Q_{A}$, and color hypercharge, $Q_{B}$. 
In terms of the usual gluon vector potential, $G$, the diagonal color gluon
states read (we denote $Q=(Q_{A},Q_{B})$)
$X=(G_{1}+iG_{2})/\sqrt{2}$ ($Q=(-1,0)$),
$Y=(G_{4}+iG_{5})/\sqrt{2}$ ($Q=(-1/2,-\sqrt{3}/2)$),
$Z=(G_{6}+iG_{7})/\sqrt{2}$ ($Q=(1/2,-\sqrt{3}/2)$).
The neutral gluons $A=G_{3}$ and $B=G_{3}$ with $Q=(0,0)$,
do not interact with monopoles at all.
Then using this basis one can easily show that the averaged over the color 
states
of the gluon and of the monopole differential cross section for $gM$
scattering reads
\beq
\frac{d\sigma_{gM}}{dq^2}=\frac{C_{A}\pi F^{2}(q^2)}{(q^2+m_{D}'^{2})^{2}}\,.
\label{eq:100}
\eeq   
One can see that similarly to scattering on thermal partons 
for monopoles the ratio of the cross sections for gluons and quarks
equals $C_{A}/C_{F}$. At $x\to 1$ the three-body cross section (\ref{eq:50}) 
reduces to the dipole cross section $\sigma_{q\bar{q}}$.
From the above consideration of the cross section for $qM$ and $gM$
scatterings we can conclude that our formula for the three-body
cross section (which has been derived for the double gluon exchanges
 \cite{NZ_sigma3})
is valid for scattering on monopoles in the limits $x\to 0,1$.
In principle for monopoles at moderate values of $x$ it may be invalid.
However, due to the fact that its variation between $x\sim 0$ and $x\sim 1$
is not very strong and the dominating region is $x\sim 0$ the 
errors due to use of (\ref{eq:50}) for monopoles cannot be significant.

For $g\to gg$ one should just replace the splitting function and $m_{q}$
by $m_{g}$ in $\epsilon^{2}$. The three-body  
cross section $\sigma_{q\bar{q}g}$  
for $g\to gg$ is replaced by the cross section for the color singlet $ggg$
state, that can be written in terms of the dipole cross section
$\sigma_{q\bar{q}}$ as
\beq
\sigma_{ggg}(\rho,x,z)=\frac{9}{8}
[\sigma_{q\bar{q}}(\rho,z)+
\sigma_{q\bar{q}}((1-x)\rho,z)
+\sigma_{q\bar{q}}(x\rho,z)]\,.
\label{eq:110}
\eeq

We use the electric Debye mass obtained in the lattice analysis 
\cite{Bielefeld_Md} 
giving $m_{D}/T$ slowly decreasing with $T$  
($m_{D}/T\approx 3.2$ at $T\sim T_{c}$, $\mu_{D}/T\approx 2.4$ at 
$T\sim 4T_{c}$). 
For the magnetic Debye mass we use predictions of the lattice simulations
of \cite{mag_M_D} that also gives decreasing with $T$ ratio $m_{D}'/T$ 
($m_{D}'/T\approx 2.8$ at $T\sim T_{c}$, and $m_{D}'/T\approx 1.2$ at 
$T\sim 4T_{c}$).  However, the energy loss is not very sensitive to the
Debye masses \cite{Z_Ecoll}.
For the quasiparticle masses of light quarks and gluon
we take $m_{q}=300$ and $m_{g}=400$ MeV  supported by 
the analysis of the lattice data \cite{LH}.
Our results are
not very sensitive to $m_{g}$, and are practically
insensitive to the value of $m_{q}$.

We used running $\alpha_s$
frozen at some value $\alpha_{s}^{fr}$ at low momenta
For gluon emission in vacuum a reasonable choice is $\alpha_{s}^{fr}\approx 0.7$
\cite{NZ_HERA,DKT}. However, in the QGP the thermal effects can suppress
the $\alpha_{s}^{fr}$, and we regard it as a free parameter.
The data on the nuclear modification factor $R_{AA}$ support
$\alpha_{s}^{fr}\sim 0.5-0.6$ for RHIC energies and
$\alpha_{s}^{fr}\sim 0.4-0.5$ for LHC energies \cite{RAA11,RAA12,RAA13}.

For the monopole density $n_{m}$ we use predictions of the recent lattice
simulations \cite{Bonati} of the $SU(3)$ gluondynamics. In \cite{Bonati}
it was obtained that the ratio $\bar{n}_{m}=n_{m}/T^{3}$ is a decreasing function of
$T$,   $\bar{n}_{m}\sim 0.9$ at $T/T_c$ slightly above 1
and $\bar{n}_{m}\sim 0.45$ at  $T/T_{c}\approx 2$. At $T/T_{c}\gsim 2$
the authors have found that 
their results may be approximated as 
$\bar{n}_{m}\approx 2A/(\ln(T/\Lambda))^{3}$
 with $A=3.66$ and $\Lambda/T_{c}=0.163$.
The difference in the monopole density for the $SU(3)$ gluondynamics
studied in \cite{Bonati} and for full QCD should not be large since the lattice 
simulations
performed in \cite{Mitr_mon} show that at $T\sim T_{c}$ the monopole density for 
the gluondynamics and for full QCD are close to each other. 
We perform the computations for point-like monopoles, i.e., for $F(q^2)=1$,
and for monopoles with a Gaussian form-factor $F(q^2)=\exp(-q^2R_{m}^{2}/6)$, 
where $R_{m}$ may be viewed as a mean square monopole radius.
Lattice results on the $r$-dependence of the monopole-(anti)monopole correlation
functions show \cite{Born_Rm1} that $R_{m}\sim 0.05-0.1$ fm. 
We are fully aware that the Gaussian form has not any theoretical
justification (especially for $F\ll 1$). But it allows one to understand how
the monopole internal structure can suppress energy loss.
We perform calculations for $R_{m}=0.1$ and $0.15$ fm.  The first value 
is consistent with the $SU(3)$ lattice simulations of \cite{Bonati}. 
The second one also may be be reasonable since
the estimation of the monopole radius via the $r$-dependence of the
monopole-(anti)monopole correlation functions 
has qualitative character rather than quantitative.

One remark is in order here on the $q(g)M$  differential cross sections.
The formulas (\ref{eq:90}), (\ref{eq:100}) do not account for the possible electric 
charge of the monopole. The lattice calculations for the $SU(2)$ gluondynamics
performed in \cite{Born_M-dyon} show that the thermal monopoles are dyons, i.e., 
besides magnetic color charge they have electric color charge as well.
It is known \cite{Urrutia} that for scattering of a charged particle (with
electric charge $q_{1e}$) on a dyon with electric and magnetic charges
$q_{2e}$ and $q_{2m}$ one should use in the Rutherford formula the 
sum $(q_{1e}q_{2m})^2+(q_{1e}q_{2e})^2$. Here only the first
(electric-magnetic) term obeys the Dirac charge quantization conditions. 
The second
(electric-electric) term is similar to that for ordinary pQCD parton-parton
interactions. One can neglect it since in the QGP the number density 
of monopoles
(dyons) is much smaller that of thermal quarks and gluons.

We have performed calculations for 
Bjorken's 1+1D expansion of the QGP \cite{Bjorken2}, 
which gives $T_{0}^{3}\tau_{0}=T^{3}\tau$. As in
our previous analyses of jet quenching \cite{RAA11,RAA12,RAA13} 
we take $\tau_{0}=0.5$ fm.
For $\tau<\tau_{0}$ we take the number density $\propto \tau$. 
We consider the situation for production of a fast parton in the central
rapidity region.
We assume that the fast parton 
produced in a hard process at $z=\tau=0$, passes through a length $L$ of an 
expanding QGP 
(our $z$-axis lies in  the impact parameter plane for $AA$ collision,
and $L$ corresponds to the transverse parton path length in the QGP
that equals the proper time $\tau$). 
We define the energy loss as
\beq
\Delta E=E\int\limits_{x_{min}}^{x_{max}} dx x\frac{dP}{dx}\,,
\label{eq:120}
\eeq
where $E$ is the initial parton energy.
Since for 
hard gluons with $x\gsim 0.5$ the jet really does not disappear, 
from the point of view of the jet quenching, a reasonable choice for the upper
limit of $x$-integration is $x_{max}=0.5$. 
For $x_{min}$ we use the value
$m_{g}/E$.

\begin{figure}[h]
\begin{center}
\epsfig{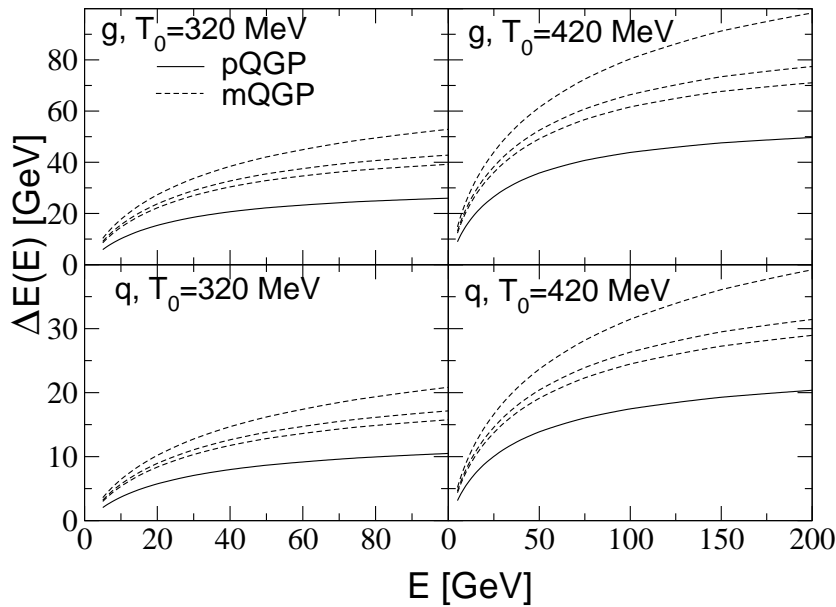}
\end{center}
\caption[.]
{
Energy dependence of the  
energy loss of gluons (upper panels) and light quarks (lower
panels) for the expanding QGP   
 with $T_{0}=320$ MeV (left) and $T_{0}=420$ MeV (right) for $L=5$ fm. 
Solid line: radiative energy loss for pQGP without monopoles; 
dashed line: radiative energy loss for mQGP 
obtained (top to bottom) for the monopole radii $R_{m}=0$, $0.1$ and $0.15$ fm.
}
\end{figure}
\begin{figure}[h]
\epsfig{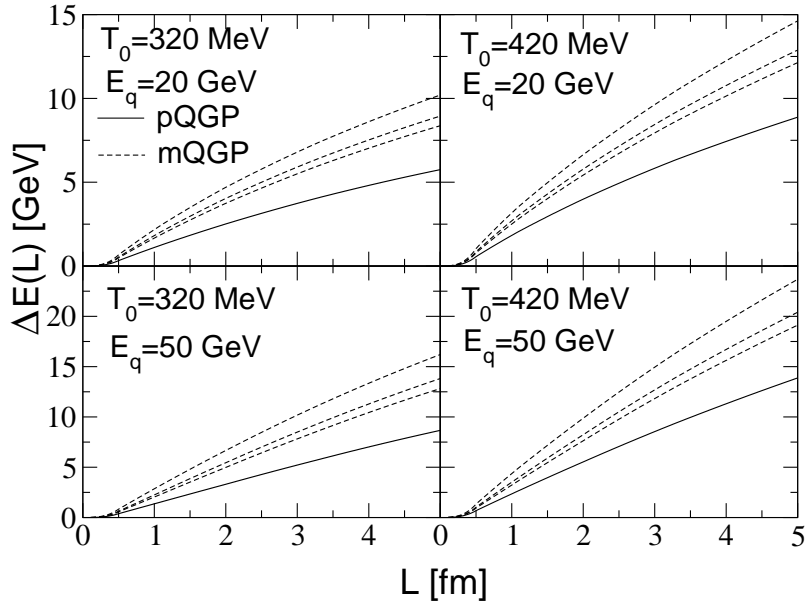}
\caption{
Radiative energy loss for quarks  with $E=20$ (upper part) and 
$50$ (lower part) GeV
vs parton path length $L$
for   the initial QGP temperature $T_{0}=320$ MeV (left) and 
$T_{0}=420$ MeV (right).
Solid line: radiative energy loss for pQGP without monopoles; 
dashed line: radiative energy loss for mQGP 
obtained (top to bottom) for the monopole radii $R_{m}=0$, $0.1$ and $0.15$ fm.
}
\end{figure}

\section{Numerical results}
We performed computations for $\alpha_{s}^{fr}=0.5$.   
We present results for 
$T_{0}=320$ MeV corresponding to central $Au+Au$ collisions
at $\sqrt{s}=0.2$ TeV, and for $T_{0}=420$ MeV corresponding to central 
$Pb+Pb$ collisions at $\sqrt{s}=2.76$ TeV (the procedure that 
leads to these values of $T_{0}$ is described in \cite{RAA13}). 
Fig.~1 shows the radiative energy loss $\Delta E$  for gluons and quarks
vs the initial parton energy $E$ for the plasma thickness $L=5$ fm 
(which approximately corresponds to the dominating parton path length 
for central $Au+Au$ and $Pb+Pb$ collisions). We show the results for 
the ordinary QGP (pQGP) and QGP with monopoles (mQGP). 
For monopoles we present the results obtained for the monopole
radii $R_{m}=0$, $0.1$ and $0.15$ fm.
From Fig.~1 one sees that monopoles may increase considerably the energy loss.
Fig.~1 shows that the enhancement factor 
$\Delta E(mQGP)/\Delta E(pQGP)$ due to
monopoles at $E\lsim 100$ GeV is $\sim 1.6-2$ 
for $R_{m}=0$ (point-like monopoles).
From the curves for $R_{m}=0.1$ and $R_{m}=0.15$ fm one can see that 
the finite-size effects can reduce the enhancement 
to $\sim 1.4-1.6$. As one could expect the effect of monopoles is 
somewhat stronger for lower $T_{0}$ since the monopole density decreases 
with temperature. However, the difference between 
the ratio $\Delta E(mQGP)/\Delta E(pQGP)$ for 
$T_{0}=320$ MeV and for $T_{0}=420$ MeV is not big.

To illustrate 
the $L$-dependence of parton energy loss
in Fig.~2 we show the results 
for radiative quark energy loss vs 
the path length $L$
for $E=20$ and $50$ GeV
for $T_{0}=320$ and $420$ MeV.
One can see that monopoles change the normalization of the curves,
but they practically do not change the form
of the $L$-dependence of $\Delta E$ (which at $L\gsim \tau_{0}$
is approximately linear).
At $L<\tau_{0}$ $\Delta E \propto L^{3}$
(since the leading $N=1$ rescattering contribution to the effective
Bethe-Heitler cross section is $\propto L$ \cite{Z_OA,Z_kin} and integration 
over the longitudinal coordinate of the scattering center gives additional 
two powers of $L$).
For the curves in Fig.~2 at $L\sim 4-5$ temperature becomes close to $T_{c}$ 
(we take $T_{c}=160$ MeV)
i.e., this region corresponds to the QCD phase transition. 
In \cite{surface}, within a purely phenomenological model
of the jet interaction with the QGP, it has been suggested  that in the 
magnetic scenario of the QGP 
the parton energy loss should be strongly enhanced in the near-$T_{c}$
region, and this should result in the surface dominance of jet energy
loss.
In terms of our curves for $\Delta E(L)$ in Fig.~2 it would mean that 
$d\Delta E/dL$ must rise steeply at $L\gsim 3$ fm.
However, our calculations 
performed with accurate treatment of induced gluon emission for realistic
lattice monopole density \cite{Bonati}, show that $\Delta E(L)$ does  not 
exhibit anything special in the near-$T_{c}$ region
for the scenario with monopoles. The curves for mQGP
may be well reproduced in the pQGP scenario by taking somewhat bigger 
$\alpha_{s}^{fr}$ (for predictions with the form-factor it is 
$\alpha_{s}^{fr}\sim 0.8$).

\section{Conclusion}
We have performed the comparison of radiative energy loss of energetic
quarks and gluons for the expanding QGP for a purely perturbative 
scenario of plasma consisting of quarks and gluons and for 
a scenario of plasma with magnetic monopoles.
Our results show that for RHIC and LHC conditions, 
for the monopole number densities in the QGP
predicted by lattice calculations, monopoles
can enhance considerably  radiative energy loss.
For point-like monopoles the enhancement factor is $\sim 1.6-2$.
Our qualitative analysis with a  Gaussian monopole form-factor 
shows that for the monopole radius $\sim 0.1$ fm, 
supported by lattice analyses, the finite-size effects can reduce
the enhancement factor to $\sim 1.4-1.6$.

Our calculations show that for RHIC and LHC conditions
the $L$-dependence of energy loss is similar to that for ordinary plasma
(except for its magnitude). It means that
monopoles do not lead to any strong surface dominance of energy loss,
as was suggested in \cite{surface}. For this reason it is difficult
to discriminate the mQGP scenario from  the pQGP one 
using data on the azimuthal anisotropy of high-$p_{T}$ hadrons
in non-central $AA$ collisions. 
Nevertheless, the effect of monopoles on the induced gluon emission
and jet quenching may be quite big and the magnetic scenario
deserves further investigation. In particular, it would be interesting
to perform a quantum analysis of the photon emission from the mQGP, 
addressed in \cite{Shuryak_m-phot} within the classical non-relativistic
formalism. This process
also can be studied consistently within the LCPI approach (in the form given
in \cite{AZ_phot}). Also, the effect of monopoles may be important
for jet quenching in $pp$ and $pA$ collisions (where the plasma temperature
is smaller than in $AA$ collisions, and monopoles may be more important), 
discussed recently in the
pQGP scenario in \cite{Z_phot,Z_RPP,Z_piter}. We leave it for future studies. 

\begin{acknowledgments}
This work is supported 
in part by the 
grant RFBR
12-02-00063-a
and the program SS-3139.2014.2.
\end{acknowledgments}


\begin{thebibliography}{99}


\bibitem{GW}
M.~Gyulassy and X.N.~Wang,
Nucl. Phys. B{\bf 420}, 583 (1994).


\bibitem{BDMPS}
R.~Baier, Y.L.~Dokshitzer, A.H.~Mueller, S.~Peign\'e, and D.~Schiff,
Nucl.\ Phys.\ B{\bf 483}, 291 (1997); {\it ibid.} B{\bf 484}, 265 (1997).
%

\bibitem{LCPI}
B.G.~Zakharov, JETP\ Lett. {\bf 63}, 952 (1996); {\em ibid}
{\bf 65}, 615 (1997);
{\bf 70}, 176 (1999);
Phys.\ Atom.\ Nucl. {\bf 61}, 838 (1998).

\bibitem{BSZ}
R.~Baier, D.~Schiff, and B.G.~Zakharov, 
Ann.\ Rev.\ Nucl.\ Part. {\bf 50}, 37 (2000), hep-ph/0002198.

\bibitem{W1}
U.A.~Wiedemann,
Nucl.\ Phys.\ A{\bf 690}, 731 (2001), hep-ph/0008241.


\bibitem{GLV1}
M.~Gyulassy, P.~L\'evai, and I.~Vitev, 
Nucl.\ Phys. B{\bf 594}, 371 (2001), hep-ph/0006010.

\bibitem{AMY}
P.~Arnold, G.D.~Moore, and L.G.~Yaffe,
JHEP {\bf 0206}, 030 (2002), hep-ph/0204343.


\bibitem{Z_Ecoll}
B.G.~Zakharov,
JETP Lett. {\bf 86}, 444 (2007), 0708.0816.

\bibitem{RAA08}
B.G.~Zakharov, JETP Lett. {\bf 88}, 781 (2008), 0811.0445.

\bibitem{RAA11}
B.G.~Zakharov,
JETP Lett. {\bf 93}, 683 (2011), 1105.2028.



\bibitem{RAA12}
B.G.~Zakharov, 
JETP Lett. {\bf 96}, 616 (2013), 1210.4148.

\bibitem{RAA13}
B.G. Zakharov,
J. Phys. G{\bf 40}, 085003  (2013), 1304.5742.

\bibitem{LH}
	
P.~L\'evai and U.~Heinz,
Phys.\ Rev.\ C{\bf 57}, 1879 (1998), hep-ph/9710463.



\bibitem{HTL}
E. Braaten and  R.D. Pisarski, 
{\sl Nucl. Phys.} B{\bf 337}, 569 (1990); B{\bf 339}, 310 (1990);\\
J. Frenkel, J.C. Taylor, {\sl Nucl. Phys.} B{\bf 334}, 199 (1990); 
B{\bf 374}, 156 (1992).


\bibitem{Shuryak_MC_mQGP} 	
J. Liao and E. Shuryak,
Phys.~Rev. C{\bf 75}, 054907  (2007), hep-ph/0611131.

\bibitem{CZ1_mQGP}
M.N. Chernodub and V.I. Zakharov,
Phys.~Rev.~Lett. {\bf 98}, 082002 (2007), hep-ph/0611228.


\bibitem{Shuryak_mQGP}
E.~Shuryak, Prog.~Part.~Nucl.~Phys. {\bf 62}, 48 (2009), 0807.3033.



\bibitem{Born_M-dyon}
V.G. Bornyakov and V.V. Braguta,
Phys.~Rev. D{\bf 84}, 074502 (2011), 1104.1063.



\bibitem{Born_Rm1}
V.G. Bornyakov and V.V. Braguta, 	
Phys.~Rev. D{\bf 85}, 014502 (2012), 1110.6308.



\bibitem{Bonati}
C. Bonati and M. D'Elia,
Nucl.~Phys. B{\bf 877}, 233 (2013), 1308.0302.


\bibitem{Mitr_mon}
V.G. Bornyakov, A.G. Kononenko, and V.K. Mitrjushkin,
arXiv:1312.4085.


\bibitem{Milton_rev}
K.A.~Milton,
Rept.~Prog.~Phys. {\bf 69}, 1637 (2006), hep-ex/0602040.


\bibitem{Shuryak_m-phot}
M.~Lublinsky, C.~Ratti, and E.~Shuryak, Phys.~Rev. D{\bf 81}, 014008 (2010),
0910.1067.


\bibitem{Gyulassy_mqgp}
J.~Xu, J.~Liao, and M.~Gyulassy, arXiv:1411.3673.

\bibitem{Z04_RAA}
B.G.~Zakharov, JETP Lett. {\bf 80}, 617 (2004), hep-ph/0410321.


\bibitem{NZ_sigma3}
N.N.~Nikolaev and B.G.~Zakharov,
Z. Phys. C{\bf 64}, 631 (1994), hep-ph/9306230.


\bibitem{Urrutia}
L.F. Urrutia, Phys.~Rev. D{\bf 18}, 3031 (1978).




\bibitem{Z_synch}
B.G. Zakharov,
JETP Lett. {\bf 88}, 475 (2008), 0809.0599.


\bibitem{Bielefeld_Md}
O.~Kaczmarek and F.~Zantow,
Phys. Rev. D{\bf 71}, 114510 (2005), hep-lat/0503017.


\bibitem{mag_M_D}
A. Nakamura, T. Saito, and S. Sakai,
Phys.~Rev. D{\bf 69}, 014506  (2004), hep-lat/0311024.


\bibitem{NZ_HERA}
N.N.~Nikolaev and B.G.~Zakharov,
Phys. Lett. B{\bf 327}, 149 (1994), hep-ph/9402209.



\bibitem{DKT}
Yu.L.~Dokshitzer, V.A.~Khoze, and S.I.~Troyan,
Phys.\ Rev. D{\bf 53}, 89 (1996), hep-ph/9506425.



\bibitem{Bjorken2}
J.D.~Bjorken, 
Phys.\ Rev. D{\bf 27}, 140 (1983).





\bibitem{Z_OA}
B.G.~Zakharov, JETP Lett. {\bf 73}, 49 (2001), hep-ph/0012360.


\bibitem{Z_kin}
B.G.~Zakharov, JETP Lett. {\bf 80}, 67 (2004), hep-ph/0406063.






\bibitem{surface}
J.~Liao and E.~Shuryak,
Phys.~Rev.~Lett. {\bf 102}, 202302  (2009), 0810.4116.


\bibitem{AZ_phot}
P. Aurenche and B.G. Zakharov,
JETP Lett. {\bf 85}, 149 (2007), hep-ph/0612343.


\bibitem{Z_phot}
B.G. Zakharov,
Phys.~Rev.~Lett. {\bf 112}, 032301 (2014), 1307.3674.


\bibitem{Z_RPP}
B.G. Zakharov,
J.~Phys. G{\bf 41}, 075008  (2014), 1311.1159.


\bibitem{Z_piter}
B.G. Zakharov,
arXiv:1412.0295.


\end{thebibliography}
\end{document}